\documentclass{ws-ijmpa}
\usepackage{amsfonts}
\usepackage{amsmath}
\usepackage{graphicx}
\usepackage{subfigure}

\newcommand{\psip}{\psi^{\prime}}
\newcommand{\psipp}{\psi(3770)}
\newcommand{\jpsi}{J/\psi}
\newcommand{\EE}{e^+e^-}

\newcommand{\pp}{\pi^+\pi^-}
\newcommand{\kk}{K^+K^-}
\newcommand{\kskl}{K^0_SK^0_L}

\newcommand{\PP}{0^-0^-}

\newcommand\Fig[1]{Fig.~\ref{#1}}
\newcommand\Table[1]{Table~\ref{#1}}

\newcommand\Eq[1]{Eq.~(\ref{#1})}

\begin{document}

\markboth{B.Q. Wang, X.H. Mo, C.Z. Yuan, Y. Ban} 
{Data taking strategy for the phase study in $\psip \to K^+K^-$}

%
\catchline{}{}{}{}{}
%

\title{Data taking strategy for the phase study in $\psip \to
  K^+K^-$}

\author{B.Q. Wang}
\address{School of Physics and State Key Laboratory of Nuclear Physics
  and Technology,\\
  Peking University, Beijing 100871, China\\
  wangbq@ihep.ac.cn}

\author{X.H. Mo}
\address{Institute of High Energy Physics, Chinese Academy of
  Sciences, \\
  Beijing 100049, China\\
  moxh@ihep.ac.cn}

\author{C.Z. Yuan}
\address{Institute of High Energy Physics, Chinese Academy of
  Sciences, \\
  Beijing 100049, China\\
  yuancz@ihep.ac.cn}

\author{Y. Ban}
\address{School of Physics and State Key Laboratory of Nuclear Physics
  and Technology,\\
  Peking University, Beijing 100871, China\\
  bany@pku.edu.cn}

\maketitle

\begin{history}
\received{Day Month Year} 
\revised{Day Month Year}
\end{history}

\begin{abstract}
  The study of the relative phase between strong and electromagnetic
  amplitudes is of great importance for understanding the dynamics of
  charmonium decays. The information of the phase can be obtained
  model-independently by fitting the scan data of some special decay
  channels, one of which is $\psi^{\prime} \to K^{+}K^{-}$. To find
  out the optimal data taking strategy for a scan experiment in the
  measurement of the phase in $\psi^{\prime} \to K^{+} K^{-}$, the
  minimization process is analyzed from a theoretical point
  of view. The result indicates that for one parameter fit, only one
  data taking point in the vicinity of a resonance peak is sufficient to
  acquire the optimal precision. Numerical results are obtained by
  fitting simulated scan data. Besides the results related to the relative
  phase between strong and electromagnetic amplitudes, the method is
  extended to analyze the fits of other resonant parameters, such as
  the mass and the total decay width of $\psi^{\prime}$.

\end{abstract}

\ccode{PACS numbers: 02.60.Pn, 13.25.Gv, 13.40.Hq}

\section{Introduction}

The charmonium hadronic decay is mainly through two processes: the
strong and the electromagnetic interactions. The relative phase
between the strong and the electromagnetic decay amplitudes is an
important parameter in understanding decay dynamics. Studies have been
carried out for many $\jpsi$ two-body decay modes:
Vector-Pseudoscalar (VP)~\cite{dm2exp,mk3exp}, Pseudoscalar-Pseudoscalar
(PP)~\cite{a00,lopez,a11}, Vector-Vector (VV)~\cite{a11} and
Nucleon-antiNucleon ($N\overline{N}$)~\cite{ann}.
These analyses reveal that there exists a relative orthogonal 
phase between the strong and the electromagnetic amplitudes in $\jpsi$
decays~\cite{dm2exp,mk3exp,a00,lopez,a11,ann,suzuki}. As to
$\psip$, there is also a theoretical argument which favors the
$\pm90^\circ$ phase~\cite{gerard}. Experimentally, some
analyses~\cite{wymppdk,wymphase,besklks1} based on limited VP and
PP data indicate that such a phase is compatible with the
data. Moreover, some efforts have been made to extend the phase study
to $\psipp$ decay phenomenologically~\cite{Wang03b,wangp06} and
experimentally~\cite{psppklks}.

The phase study can provide valuable clue for exploring the relation
between the strong and the electromagnetic interactions. Now with the
upgraded accelerator BEPCII~\cite{bepcii} and detector
BESIII~\cite{besiii}, a luminosity of $6.5\times
10^{32}$cm$^{-2}$s$^{-1}$ has achieved, which is the highest
luminosity in $\tau$-charm energy region ever existed. 226 M
$J/\psi$ events, 106 M $\psi^{\prime}$ events, and 2.9 fb$^{-1}$
$\psi(3770)$ data have been collected~\cite{bes3data}, even more
colossal data are to be collected in the forthcoming years, which
gives a great opportunity to determine the phase between the strong
and the electromagnetic amplitudes with unprecedented statistical
precision.

However, examining the existing determination of the relative phase,
since the data are merely taken at one or two energy points, we find
most of studies are model-dependent. A typical model assumption is
the SU(3) symmetry in charmonium decays which supply additional
constraint on the electromagnetic decay amplitudes in charmonium
decays into similar final states such as VP, PP and so on. Now with
a high luminosity accelerator, it is possible to measure the phase
model-independently by scanning the cross sections in the vicinity 
of the resonance. As the strength of the resonance decays varies with 
energy, the precision of the phase measurement depends on the data 
taking energy when the total data taking time is fixed.
Therefore, the optimization study for the data taking strategy
is of great importance in order to obtain the most precise results
with the limited luminosity (equivalently within the limited data
taking time).

Without losing generality, we focus on the mode of $\psip$ decays to
$\kk$ final state. Because, as will be shown in the next section, this
decay mode can accommodate a comparatively simple parametrization form
which is of great benefit to extract the relative phase.

As far as the optimization of data taking strategy is concerned,
sampling simulation technique was adopted for optimizing the
$\tau$ mass measurement~\cite{taustat,taustat2}.  An
interesting conclusion from the study is that for one parameter fit,
data at only one energy point is enough to acquire the best precision.
The breakthrough of this monograph lies in that the minimization
process is analyzed in detail from a theoretical point of view, which
leads to the same conclusion as that of $\tau$ mass measurement. As
a cross check, numerical results are obtained by  fitting simulated
$\psi^{\prime}$ scan data. Moreover, this method is extended to extract
other resonance parameters, such as the mass and
the total decay width of $\psi^{\prime}$.

\section{Theoretical Framework}

For $\psip \to \PP$,
the $\pp$ channel is through electromagnetic decays, the $\kskl$
through SU(3) breaking strong decays, and the $\kk$ through
both. Therefore, the $\psip \to \kk$ decay is the only process which
can be used to study the phase between strong and electromagnetic
interactions in an energy scan experiment. Taken into account the continuum
process, the decay amplitude of this mode is parametrized
as~\cite{wymppdk,wymphase,Wang03}:

\begin{equation}
A_{K^+K^-}=E_c+E+\frac{\sqrt{3}}{2}M,
\end{equation}
where $E_c$ is the continuum amplitude,
 $E$ the electromagnetic amplitude, and $\frac{\sqrt{3}}{2}M$
 the SU(3) breaking strong amplitude. They can be
expressed explicitly as

\begin{equation}
E_c \propto \frac{1}{s},\quad E \propto \frac{1}{s}B(s), \quad
\frac{\sqrt{3}}{2}M \propto Ce^{i\phi}\frac{1}{s}B(s),
\end{equation}
where the real parameters $\phi$ and $C$ are the relative phase and
the relative strength between the strong and the electromagnetic
amplitudes, and $B(s)$ is defined as~\cite{wymppdk}

\begin{equation}
B(s)=\frac{3\sqrt{s}\Gamma_{ee}/\alpha}{s-M^2+iM\Gamma_t},
\end{equation}
where $\sqrt{s}$ is the center of mass energy, $\alpha$ is the QED
fine structure constant, $M$ and $\Gamma_t$ are the mass and total
width of $\psip$, $\Gamma_{ee}$ is the partial width of $\psip\to \EE$.

The Born order cross section for this channel reads

\begin{equation}
\begin{split}
    \sigma^{Born}_{K^+K^-} = \frac{4\pi\alpha^2}{s^{3/2}}\left[1 +
    2\mathfrak{R}(C_{\phi}B(s)) + |C_{\phi}B(s)|^2 \right] \\ \times
    |F_{K^+K^-}(s)|^2 P_{K^+K^-}(s),
  \end{split}
\end{equation}
where $C_{\phi} = 1 + Ce^{i\phi}$; $F_{K^+K^-}(s)$ is the form factor,
which is usually written as $F_{K^+K^-}(s) = f_{K^+K^-}/s$ with
$f_{K^+K^-}$ being a constant; $P_{K^+K^-}(s) =
\frac{2}{3s}~ q_K^3$ is the phase space factor, with
$$
    q_K^2 = E_K^2 - m_K^2 = \frac{s}{4} - m_K^2,
$$
where $q_K$ is the momentum of $K^+$ or $K^-$, $m_K$ is the nominal
mass of $K$ meson.

In actual experiment, the effect of Initial State Radiation (ISR)
is considered through an integral~\cite{Kuraev85}

\begin{equation}
    \sigma_{r.c.}(s) = \int_0^{X_f} dx F(x,s)
    \sigma_{Born}(s(1-x)),
	\label{eq-isr}
\end{equation}
where $F(x,s)$ is the structure function which can be calculated to an
accuracy of 0.1\%~\cite{Kuraev85,Altarelli,Berends}.

In addition, another important experimental effect, the energy spread
of $e^+$ and $e^-$ must also be taken into consideration. Finally, the
experimentally observed cross section is expressed as~\cite{bk:Lee,bk:Wille}

\begin{equation}
\sigma_{exp}(\sqrt{s}) = \int_0^{\infty} d\sqrt{s'}
\sigma_{r.c.}(\sqrt{s'}) G(\sqrt{s'},\sqrt{s}),
\label{eq-exp}
\end{equation}
where $G(\sqrt{s'}, \sqrt{s})$ is a Gaussian distribution
$$
G(\sqrt{s'}, \sqrt{s})=\frac{1}{\sqrt{2\pi}\Delta}
e^{-\frac{(\sqrt{s'} - \sqrt{s})^2} {2\Delta^2}}.
$$
Here $\Delta$ indicates the energy spread of the collision beams.

Some parameter values for the numerical calculation in the following
sections are articulated in Table~\ref{tab:parametervalue}.

\begin{table}[ph]
  \tbl{Some parameter values for numerical calculation. The quantity with $\star$
    will be set as a free fitting parameter for the corresponding study. }
  {\begin{tabular}{@{}lrl@{}}
\toprule
        Quantity    & numerical value & Remark  \\
\colrule
    $\star M$       & 3.68609 GeV     & Ref.~\cite{pdg10}  \\
    $\star \Gamma_t$& 304 keV         & Ref.~\cite{pdg10}  \\
      $\Gamma_{ee}$ & 2.35 keV        & Ref.~\cite{pdg10}  \\
      $m_K$         & 493.677 MeV     & Ref.~\cite{pdg10}  \\
      $f_{\kk}$     &  0.9 GeV$^2$    & Ref.~\cite{pedler}  \\
       $\Delta$     &  1.3 MeV        & Ref.~\cite{moxh2002}  \\
   $\star \phi$     & 90$^{\circ}$    & Ref.~\cite{cleo06}  \\
         C          & 2.5            & Ref.~\cite{cleo06} \\
\botrule
  \end{tabular}
  \label{tab:parametervalue}}
\end{table}

\section{Minimization Analysis}
\label{minimize}

For a scan experiment, several points, say totally $N_{pt}$ points,
need to be taken in a vicinity of a resonance (in this monograph
the $\psip$). The estimator is usually constructed as~\cite{bk:zys}:

\begin{equation}
    \chi^2=\sum_{i=1}^{N_{pt}}
    \frac{(N_i^{obs}-L_i\sigma_i\varepsilon_i)^2} {(\Delta N_i^{obs})^2},
    \label{chi2}
\end{equation}
where $N_i^{obs}$ and $\Delta N_i^{obs}=\sqrt{N_i^{obs}}$ are the
observed number of events and its error at the $i$-th point, $L_i$ the
corresponding luminosity, $\varepsilon_i$ the selection efficiency,
and $\sigma_i$ the theoretical cross section that is $\sigma_{exp}$ in
\Eq{eq-exp}. The fitting parameters (relative phase, strength, etc.)
are contained in $\sigma_i$, and these parameters and the
corresponding errors can be extracted by minimizing the $\chi^2$
function defined in \Eq{chi2}. In the following analyses, only
concerned is one free fitting parameter, the relative phase between
strong and electromagnetic amplitudes, that is, $\phi$.

If we denote the observed cross section measured at energy point $i$ as
$\sigma_i^{obs}$, and rewrite
\begin{equation}
  N_i^{obs} = L_i \sigma_i^{obs} \varepsilon_i~,
\end{equation}
\Eq{chi2} can be recast as
\begin{equation}
  f = \sum_i \frac{L_i \varepsilon_i} {\sigma_i^{obs}}
  (\sigma_i^{obs} - \sigma_i)^2 = L_0 \varepsilon \sum_i \frac{x_i}
  {\sigma_i^{obs}} (\sigma_i^{obs}-\sigma_i)^2~.
\label{f}
\end{equation}
Here $\chi^2$ is replaced with $f$ for simplicity, and the following relations
are utilized :
\begin{equation}
L_i=x_i L_0~, \qquad \sum_i x_i =1~,
\end{equation}
where $L_0$ is the total luminosity (corresponding to the finite total
data taking time) and $x_i$ is the fraction of luminosity at the
$i$-th energy point.  Moreover, $\varepsilon_i$ is supposed to be the
same at all points ($\varepsilon=50\%$ is used for numerical
calculation), which is a fairly good approximation for the scan of
narrow resonances, such as $J/\psi$ and $\psi^{\prime}$.

In the light of \Eq{f}, the first and second order derivatives of the
function $f$ to $\phi$ can be derived as

\begin{equation}
  \frac{\partial f}{\partial \phi} = L_0 \varepsilon \sum_i
  \frac{x_i}{\sigma_i^{obs}}  2(\sigma_i^{obs}-\sigma_i)(-\frac{\partial
    \sigma_i} {\partial \phi}),
\label{f-deriv1}
\end{equation}

\begin{equation}
  \frac{\partial^2 f}{\partial \phi^2}
  = 2L_0\varepsilon \sum_i \frac{x_i}{\sigma_i^{obs}} \left[ \left(
      \frac{\partial \sigma_i}{\partial \phi} \right)^2 -
    (\sigma_i^{obs} - \sigma_i)\left( \frac{\partial^2
        \sigma_i}{\partial \phi^2} \right) \right] ~.
\label{f-deriv2}
\end{equation}

\begin{figure}
\centering
\subfigure[]{\includegraphics[angle=-90,width=0.4\textwidth]{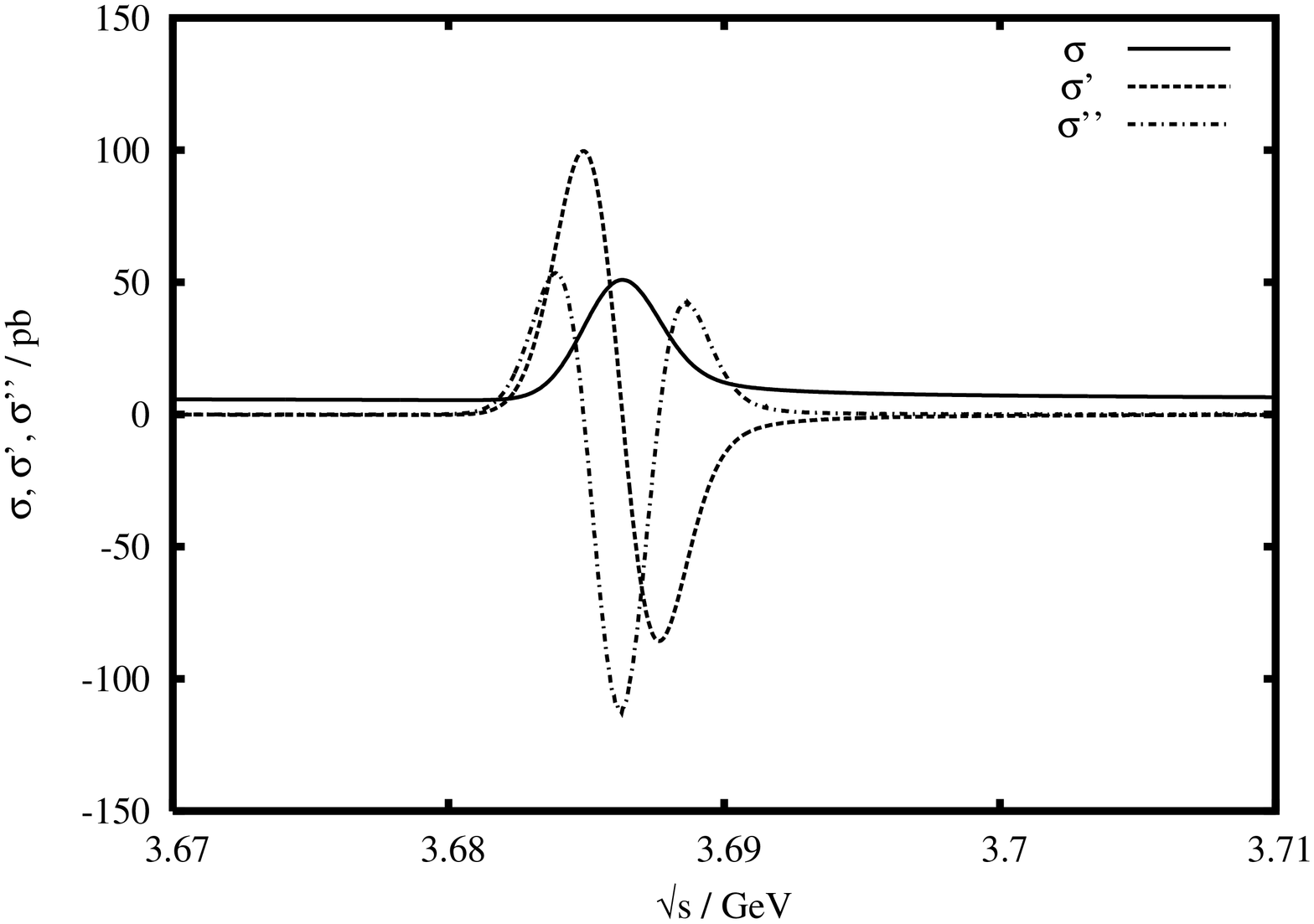}}
\subfigure[]{\includegraphics[angle=-90, width=0.4\textwidth]{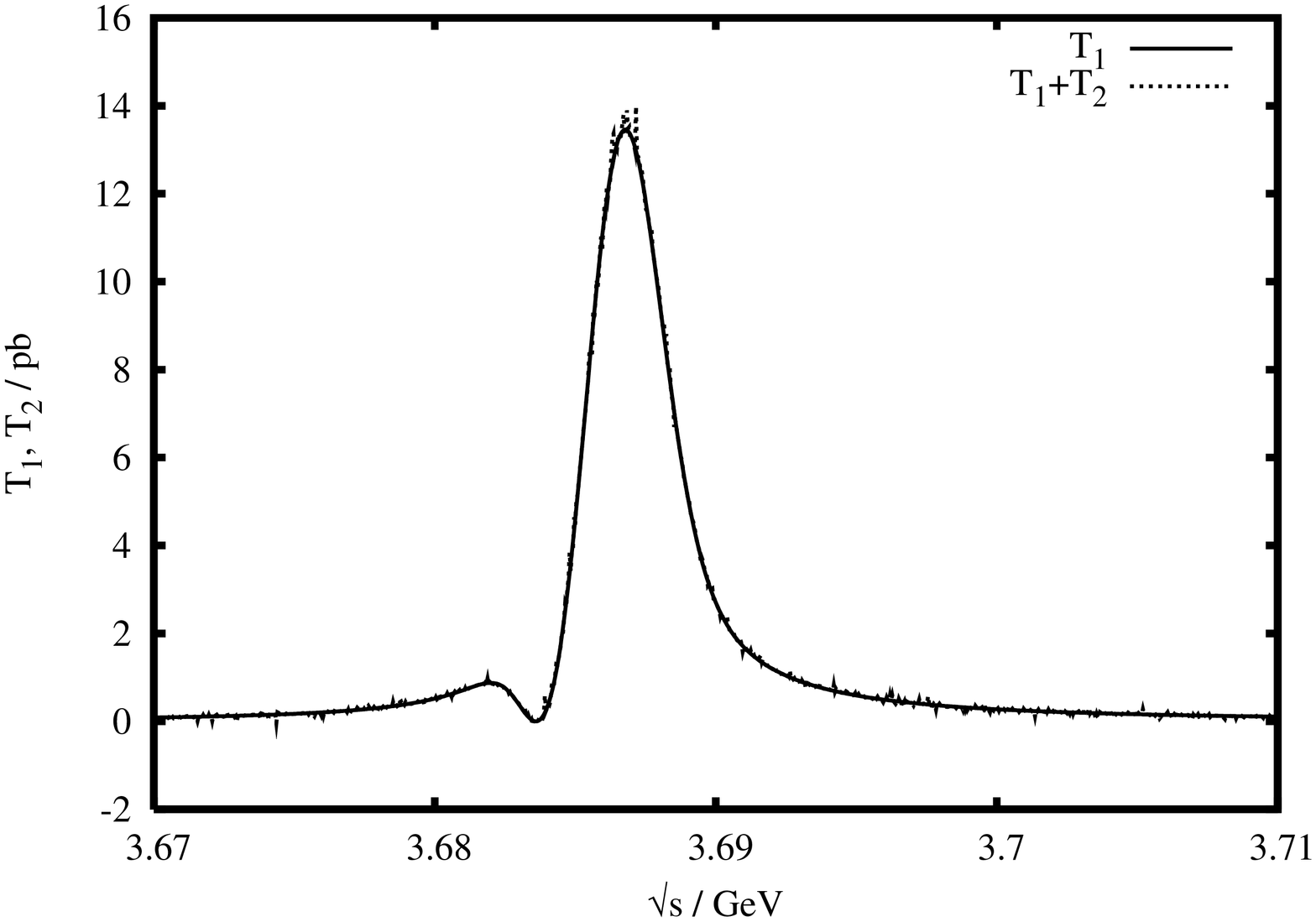}}
\caption{(a) Comparison between the cross section and its first and
  second order derivative. The solid line shows the cross section
  ($\sigma$). The dashed line shows the first order derivative
  ($\sigma'$). The dotted-dashed line shows the second order
  derivative ($\sigma''$). (b) Comparison between first and second
  term inside the bracket in \Eq{f-deriv2}. The solid line only
  shows the value of the first term (T$_1$); the dotted line contains
  contributions both from the first term and the second term (T$_1$ +
  T$_2$). Values in Table~\ref{tab:parametervalue} are taken in making 
  these plots.}
  \label{fig:term-comp}
\end{figure}

The experimentally concerned cross section functions are generally
smooth enough, which can be approximated by polynomial
functions. Therefore, the first and second order derivatives of these
functions are also smooth enough (refer to \Fig{fig:term-comp} 
when the parameters take the values in Table~\ref{tab:parametervalue}).
Under such case, we argue that the second term in \Eq{f-deriv2} could
be neglected. When the fitting process finishes, $\sigma_i$ in
\Eq{f-deriv1} and \Eq{f-deriv2} can be considered as the true value
of the cross section at energy point $i$. As we assumed previously,
$\sigma_i^{obs}$ is the experimentally measured cross section at
energy point $i$, then ($\sigma_i^{obs} - \sigma_i$) could be
considered as a random variable which satisfy a Gaussian distribution
with mean as 0 and deviation as $\Delta \sigma_i^{obs}$ (the error of
$\sigma_i^{obs}$). As a conservative estimation, we assume the
relative error of cross section measurement is 10\%, which means
$\Delta \sigma_i^{obs} = 0.1\cdot \sigma_i^{obs}$. The
expectation of the second term inside the sum in \Eq{f-deriv2} could
be calculated by using sampling method. The comparison of these two
terms are shown in \Fig{fig:term-comp}, from which the second term is
quite small compared with the first one, therefore its effect can be
neglected. Now \Eq{f-deriv2} becomes

\begin{equation}
  \left.\frac{\partial^2 f}{\partial \phi^2} \right|_{\phi=\phi^*} = 2L_0
  \varepsilon \sum_i \frac{x_i} {\sigma_i^*}\left. \left(\frac{\partial \sigma_i}
  {\partial \phi}\right)^2  \right|_{\phi=\phi^*},
\label{f2}
\end{equation}
where $\phi^{*}$ is the fitting result of the relative phase and
$\sigma_i^{*}$ is the theoretical cross section at $\phi=\phi^{*}$.

The fitting error of $\phi$ can be evaluated as~\cite{bk:zys}
\begin{equation}
E(\phi) = \sqrt{2} \cdot \left. \left( \frac{\partial^2 f}{\partial \phi^2}\right)^{-1/2}\right|_{\phi=\phi^*}~.
\label{phi-err}
\end{equation}
According to \Eq{phi-err}, the maximum of the second order derivative
of fitting function yields the minimum of fitting error. Define a new
function $g$ as
\begin{equation}
  \label{eq:g}
g_i \equiv \frac{1}{\sigma_i^*}\left.\left(\frac{\partial \sigma_i}{\partial \phi}\right)^2 \right|_{\phi=\phi^*}~,
\end{equation}
where the subscript $i$ denotes the value of $g$ at the $i$-th energy
point. Then \Eq{f2} becomes

\begin{equation}
  \label{eq:f2-g}
 \left. \frac{\partial^2 f}{\partial \phi^2}\right|_{\phi=\phi^*} = 2L_0
  \varepsilon \sum_i x_i g_i .
\end{equation}

Notice that $\sum_i x_i=1$, it is readily to obtain the following inequalities
\begin{equation}
	g_{min} = (\sum_i x_i) g_{min} \le \sum_i x_i g_i \le (\sum_i x_i) g_{max} = g_{max},
\end{equation}
where $g_{min}$ ($g_{max}$) is the minimum (maximum) value of $g$
within the energy region concerned. To get maximum $\sum_i x_i g_i$,
only one data taking point is sufficient and it should be located at
the energy point which renders $g$ maximum.

\section{Numerical results}

To reinforce the preceding conclusion, the simulated scan data are fit
to get numerical results. In this procedure, great many times of
fitting need to be performed, where the large number of calculations
must be carried out for the observed cross section. Unfortunately, two
nested integrations of the observed cross section, which take into
account both the ISR correction and beam energy spread effect, take so
much time that any actual optimization fitting becomes impractical. In
a recent study~\cite{wangbq10}, using the generalized linear
regression approach, a complex energy-dependent factor is approximated
by a linear function of energy. Taken advantage of this
simplification, the integration of ISR correction can be performed and
an analytical expression with accuracy at the level of 1\% is
obtained. Then, the original two-fold integral is simplified into a
one-fold integral, which reduces the total computing time by two
orders of magnitude. In the following studies, the simplified observed
cross section formulas are adopted to acquire all numerical results.

\subsection{Relative Phase}
\label{sec:par:phase}

\begin{figure}
  \centering
  \includegraphics[angle=-90, width=0.8\textwidth]{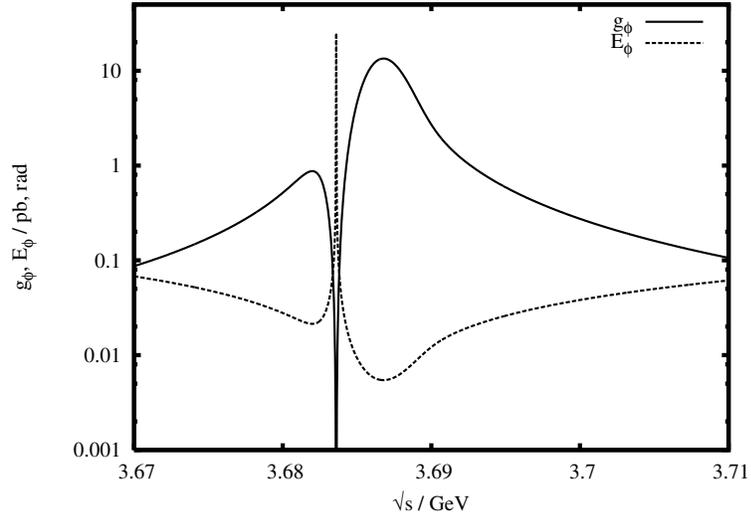}
\caption{The distributions of $g_{\phi}$ and $E_{\phi}$ for $\phi^*=90^{\circ}$.}
\label{fig-g-90}
\end{figure}

Considering the parameter to be analyzed is the relative phase between
strong and electromagnetic amplitude of $\psip$ decay, the
distribution of $g_{\phi}$ and the fitting error $E_{\phi}$ on energy region when
$\phi^*=90^{\circ}$ is shown in \Fig{fig-g-90},
according to which, at the energy point 3.6868 GeV, the value of
function $g_{\phi}$ reaches its maximum while $E_{\phi}$ reaches its minimum.
In the vicinity of 3.684 GeV, the $g_{\phi}$ value is very small and the
corresponding $E_{\phi}$ is quite large. So this point (3.684 GeV)
should be avoided in the scan experiment\footnote{To validate this result,
the sampling technique is used to check the data taking point distribution
and the fitting error. Details about the sampling technique can be found
in Appendix.}.

\begin{figure}
\centering
\subfigure[]{\includegraphics[angle=-90,width=0.4\textwidth]{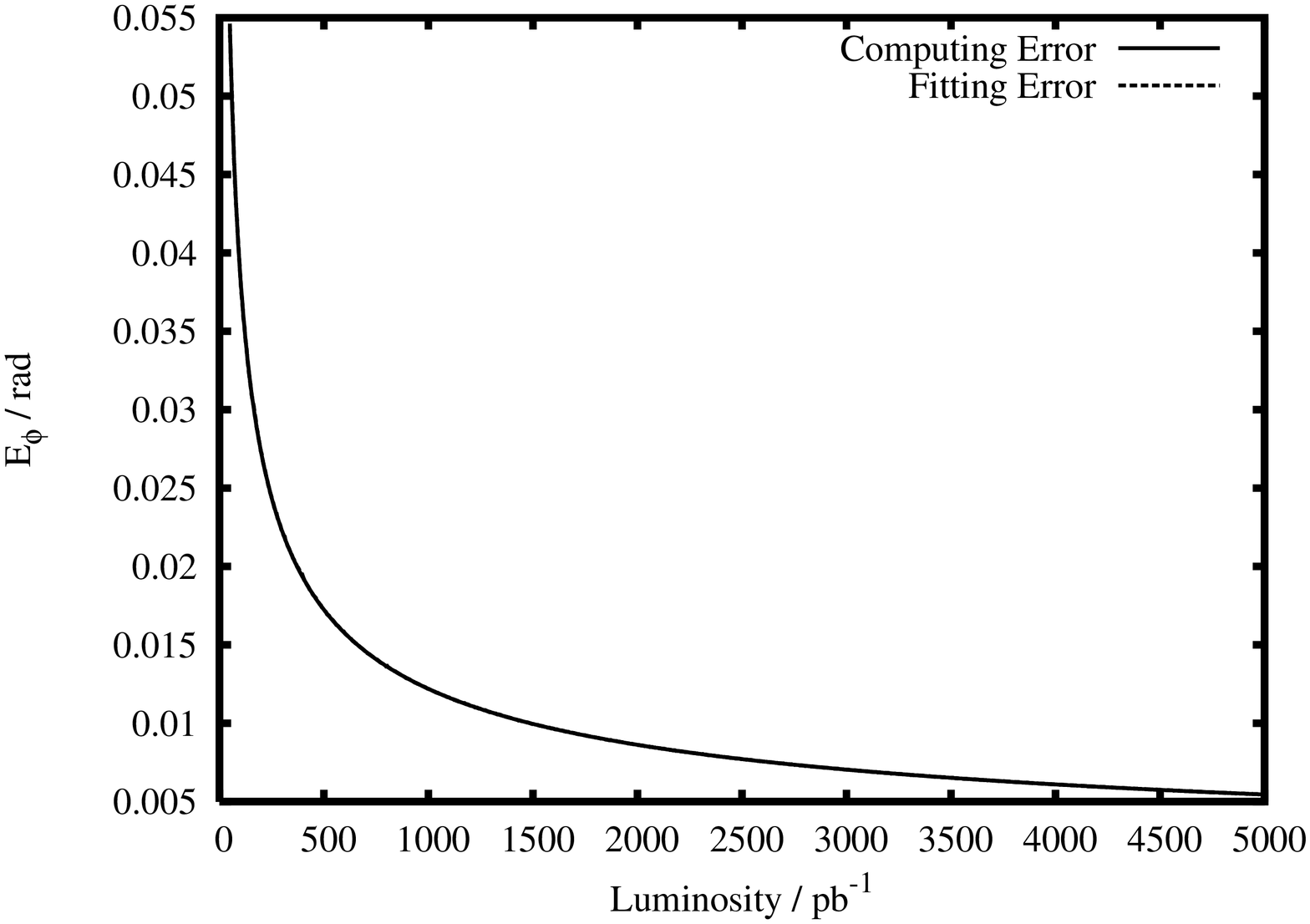}}
\subfigure[]{\includegraphics[angle=-90,width=0.4\textwidth]{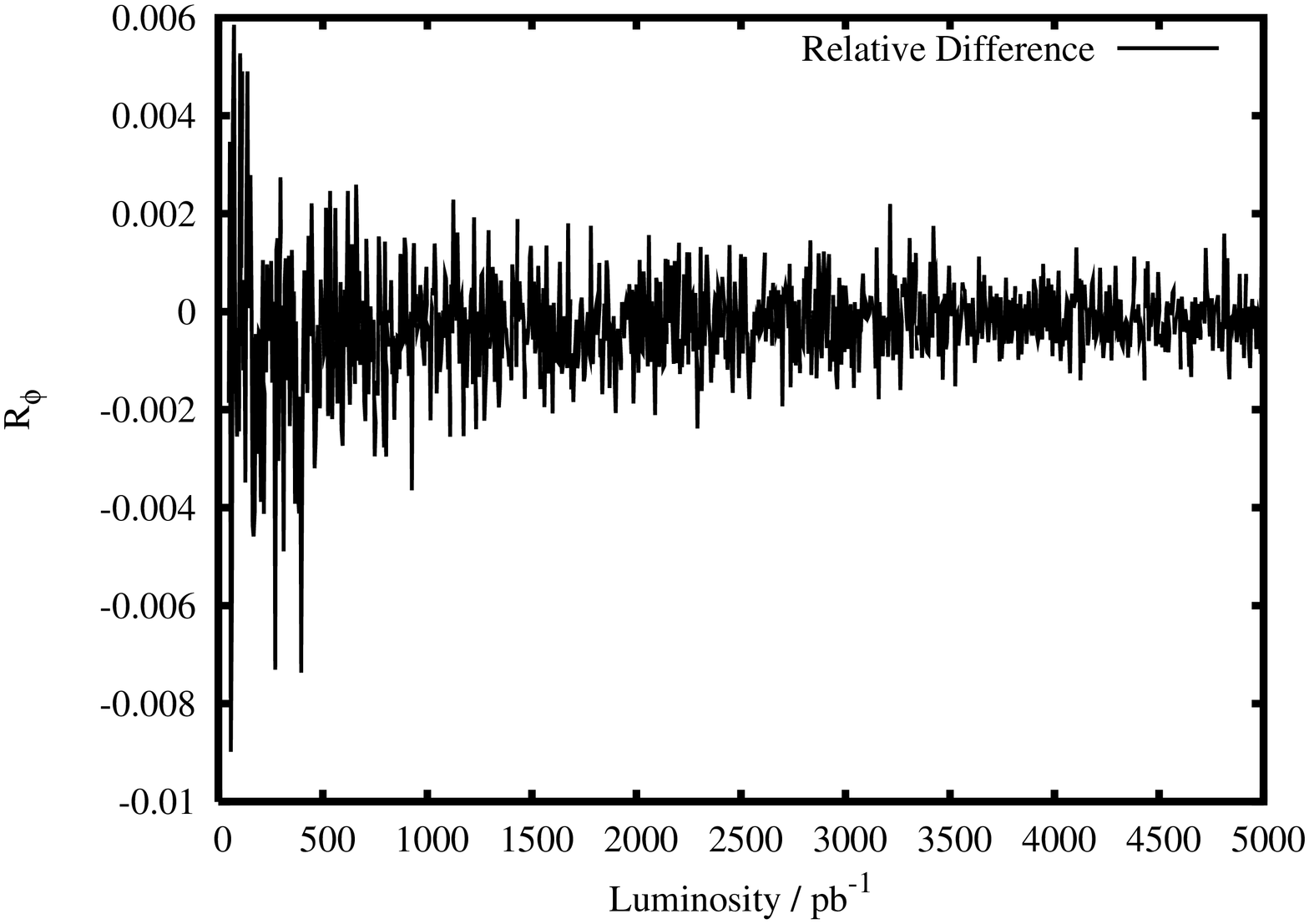}}
\caption{(a) Comparison of the equivalent error for different
  luminosity by fixing the energy point at 3.6868 GeV; (b) Relative
  difference $R_{\phi}$ at 3.6868 GeV for different luminosity.}
\label{scan-err-comp}
\end{figure}

By fixing the energy point to 3.6868 GeV, the error obtained
from fitting and computed by \Eq{phi-err} is shown in \Fig{scan-err-comp}.
Just as expected, the higher the luminosity, the smaller the error.
Moreover, the fitting and computing values of error are so consist
with each other that it is hardly to distinguish them in \Fig{scan-err-comp} (a).
To exhibit the details, the relative difference of $E_{\phi}$, that is,
$$
R_{\phi}=\frac{ E_{\phi}\text{(computing)}-E_{\phi}\text{(fitting)} }
{E_{\phi}\text{(computing)}}
$$
is shown in \Fig{scan-err-comp} (b).

If the relative phase $\phi$ variates, the optimal position of
energy will change correspondingly. \Table{tab:phase:opt} lists
the optimal values of energy position for some special phase angles.
According to these information, the values of the optimal
energy points arrange from 3.686 GeV $\sim$ 3.687 GeV, nearly
within the scope of 1 MeV.

\begin{table}[htbp]
  \tbl{The optimal data taking position for different relative phase $\phi$}
  {\begin{tabular}{@{}cc@{}}
\toprule
    $\phi$($^{\circ}$) & Optimal point (GeV) \\
\colrule
    0 & 3.68604 \\
    45 & 3.68700 \\
    90 & 3.68680 \\
    135 & 3.68706 \\
    180 & 3.68648 \\
    270 & 3.68672 \\
\botrule
  \end{tabular}
  \label{tab:phase:opt}}
\end{table}
\subsection{Other parameters}
\label{sec:par:other}

The minimization analysis discussed in Section~\ref{minimize} is
applicable to any parameter we are concerned with. As long as the
variable $\phi$ is replaced with the parameter to be analyzed, all
aforementioned deductions are valid. In study that followed,
we perform the optimization for the two interested resonance parameters,
mass and total width.

\subsubsection{$\psip$ Mass}

The similar analyses are performed for the mass of $\psip$ resonance,
and results are displayed in \Fig{fig:err:gamt}, where four curves
corresponding to different relative phases, $\phi = 0^{\circ},
90^{\circ}, 180^{\circ},$ and $270^{\circ}$.  There are two new
features for the optimization of mass parameter. Firstly, the energy
position for smallest $E_M$ is at 3.6845 GeV for $\phi$ = $0^{\circ}$,
$90^{\circ}$, and $270^{\circ}$; but at 3.6874 GeV for $\phi =
180^{\circ}$.  Secondly, two energy positions should be avoided due to
larger values of $E_M$. One is around 3.686 GeV, the other is near
3.68 GeV for $\phi = 0^{\circ}, 90^{\circ},$ and $270^{\circ}$; while
near 3.69 GeV for $\phi = 180^{\circ}$.

\begin{figure}
\begin{minipage}{0.49\linewidth}
\subfigure[]{\includegraphics[angle=-90,width=1.0\textwidth]{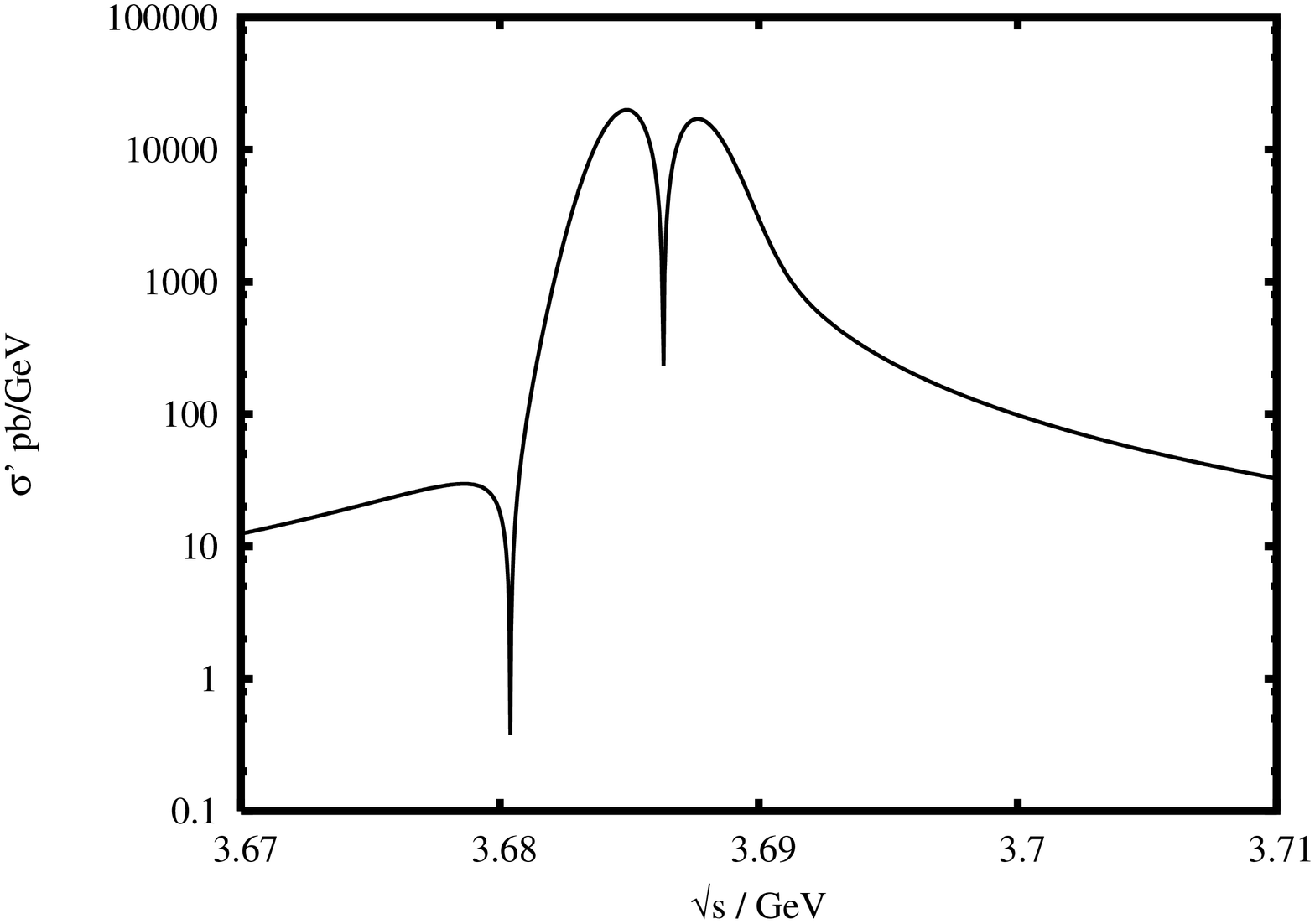}}
\subfigure[]{\includegraphics[angle=-90,width=1.0\textwidth]{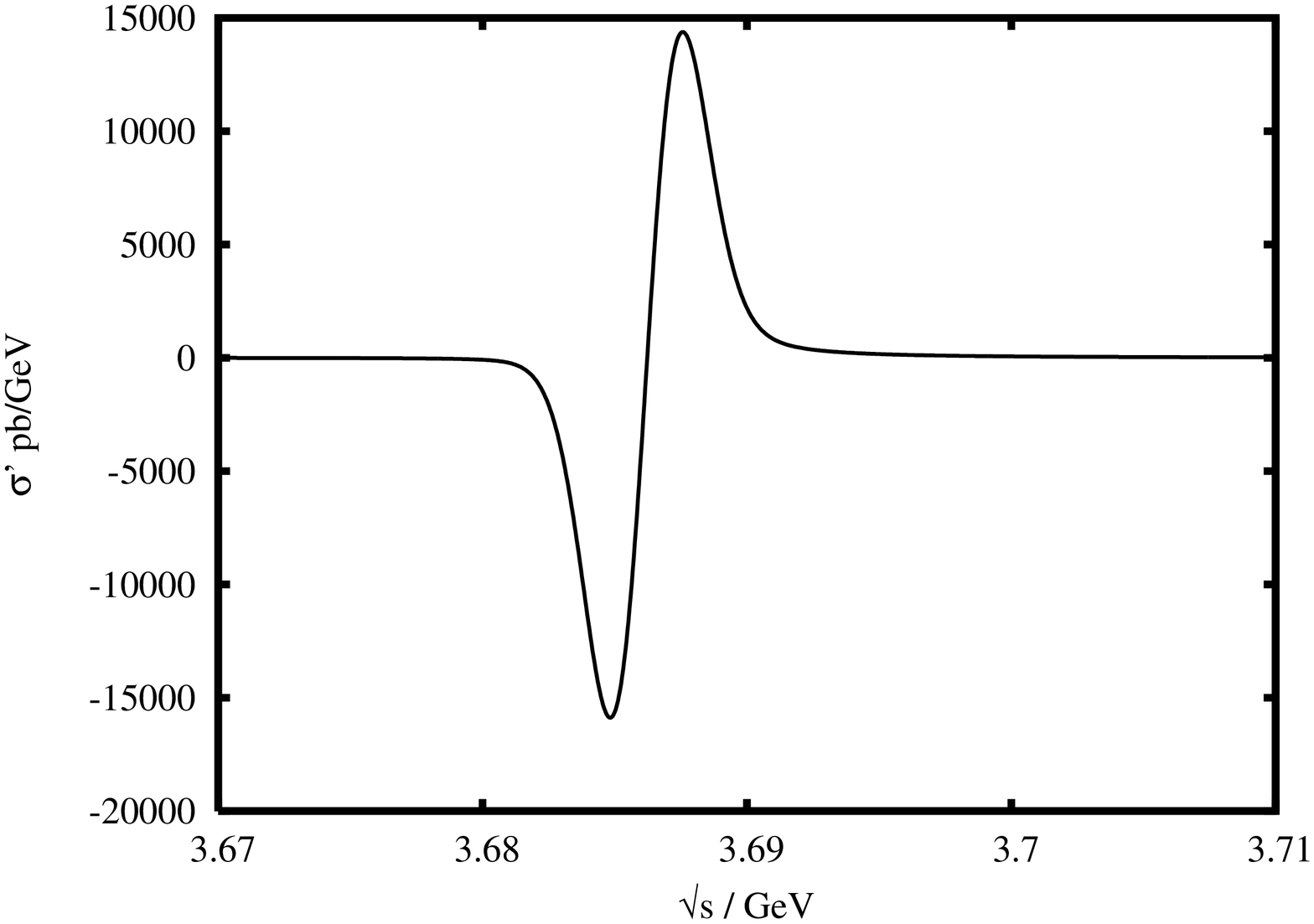}}
\subfigure[]{\includegraphics[angle=-90,width=1.0\textwidth]{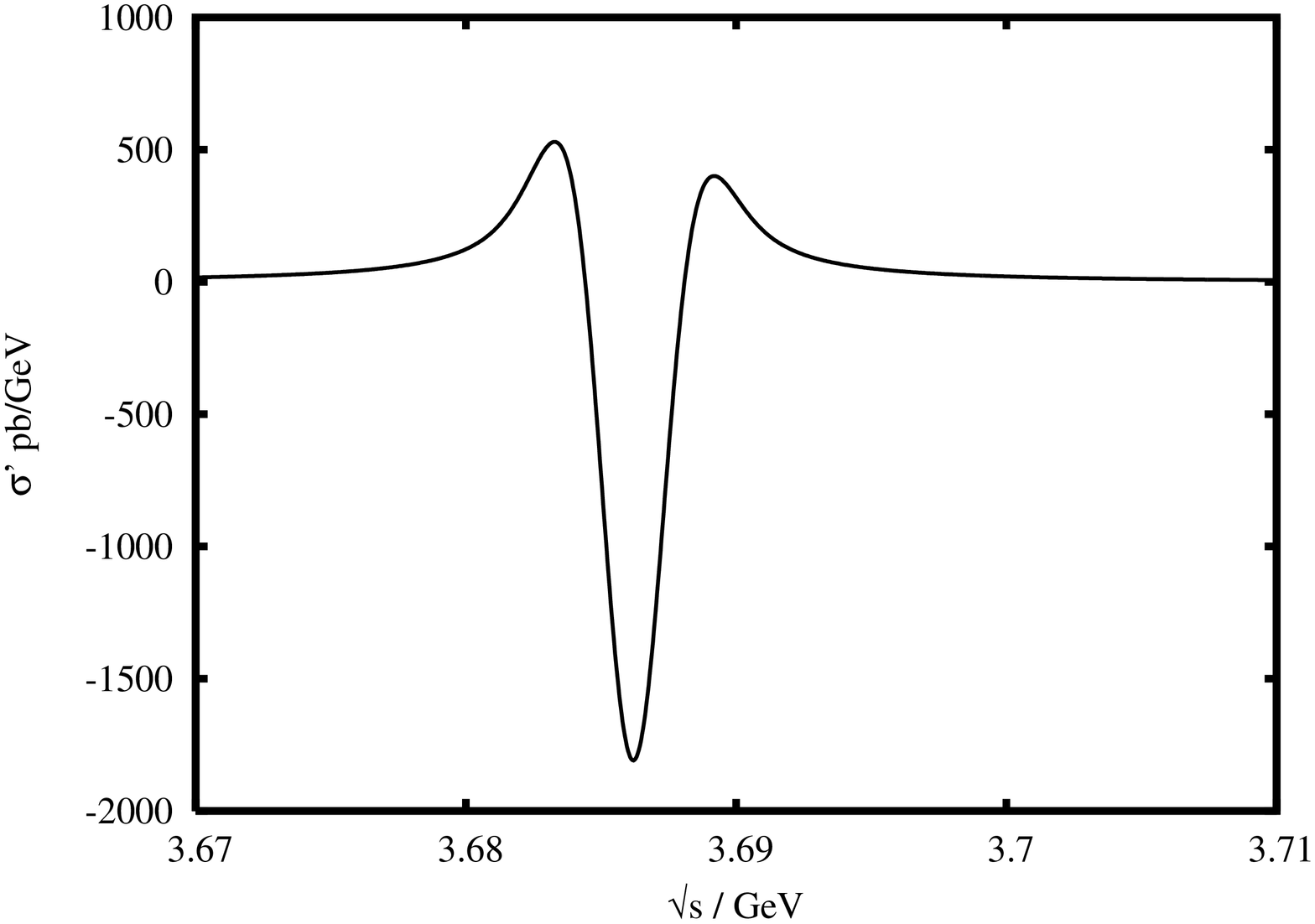}}
\subfigure[]{\includegraphics[angle=-90,width=1.0\textwidth]{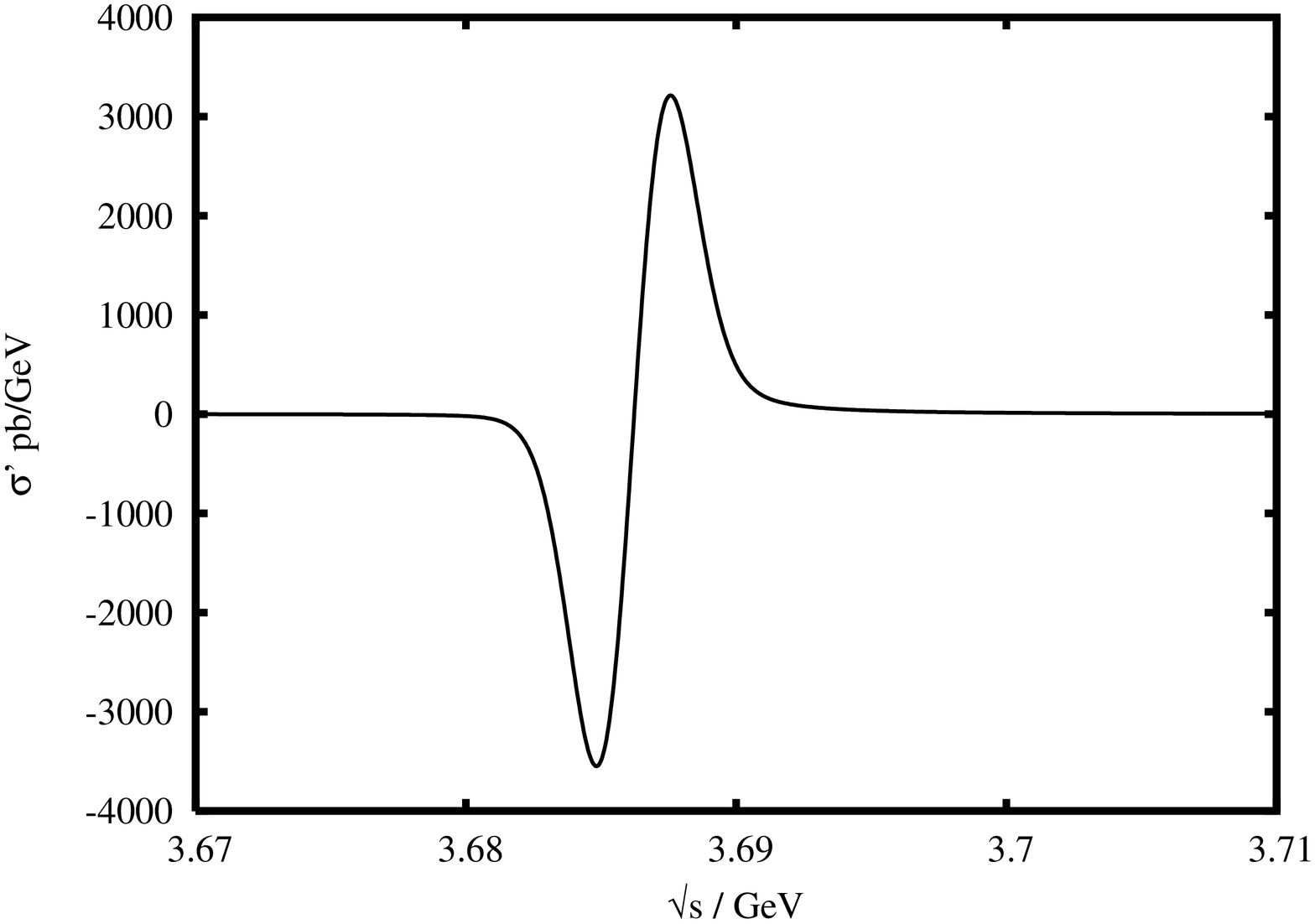}}
\end{minipage}                                
\begin{minipage}{0.49\linewidth}              
\subfigure[]{\includegraphics[angle=-90,width=1.0\textwidth]{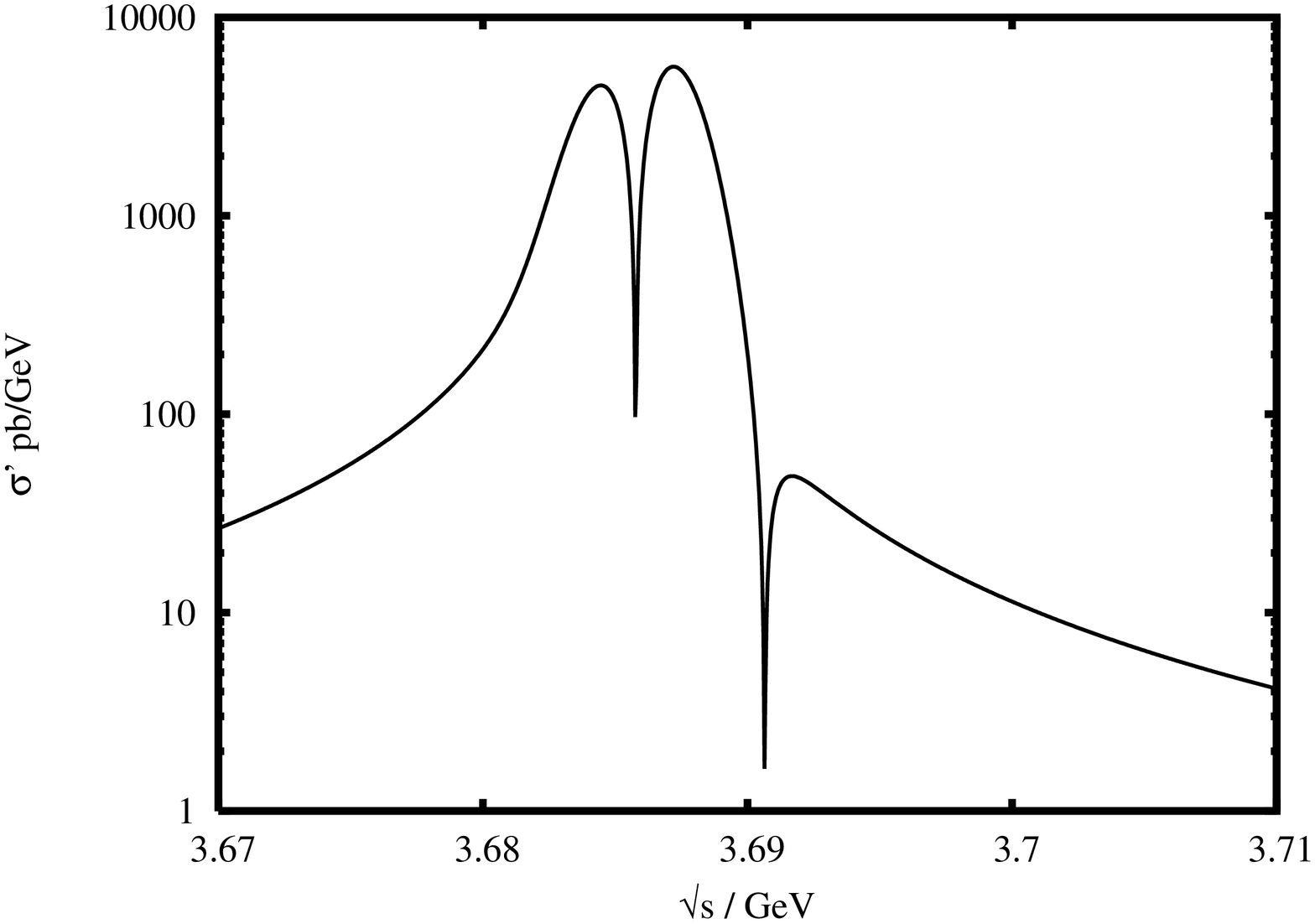}}
\subfigure[]{\includegraphics[angle=-90,width=1.0\textwidth]{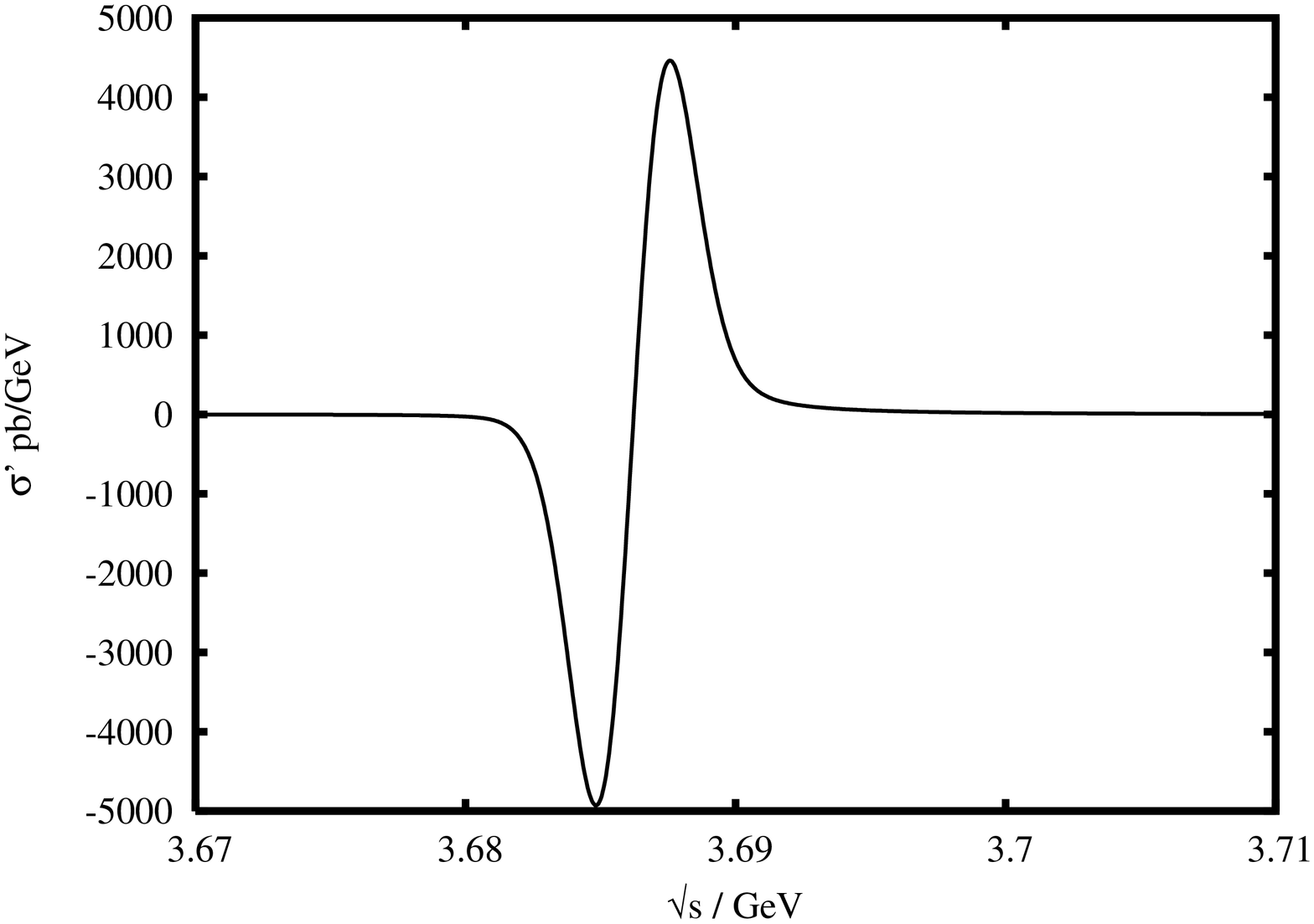}}
\subfigure[]{\includegraphics[angle=-90,width=1.0\textwidth]{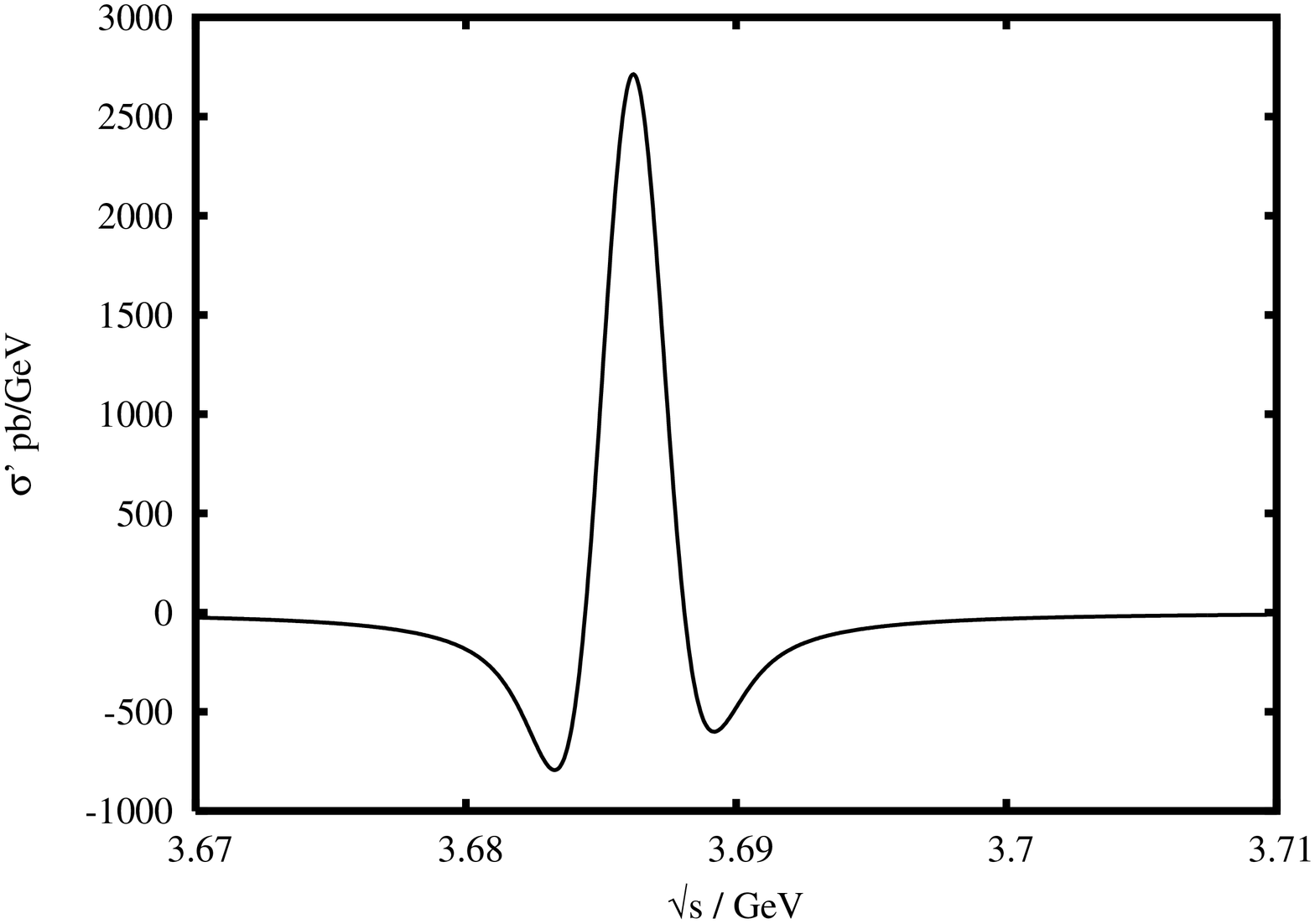}}
\subfigure[]{\includegraphics[angle=-90,width=1.0\textwidth]{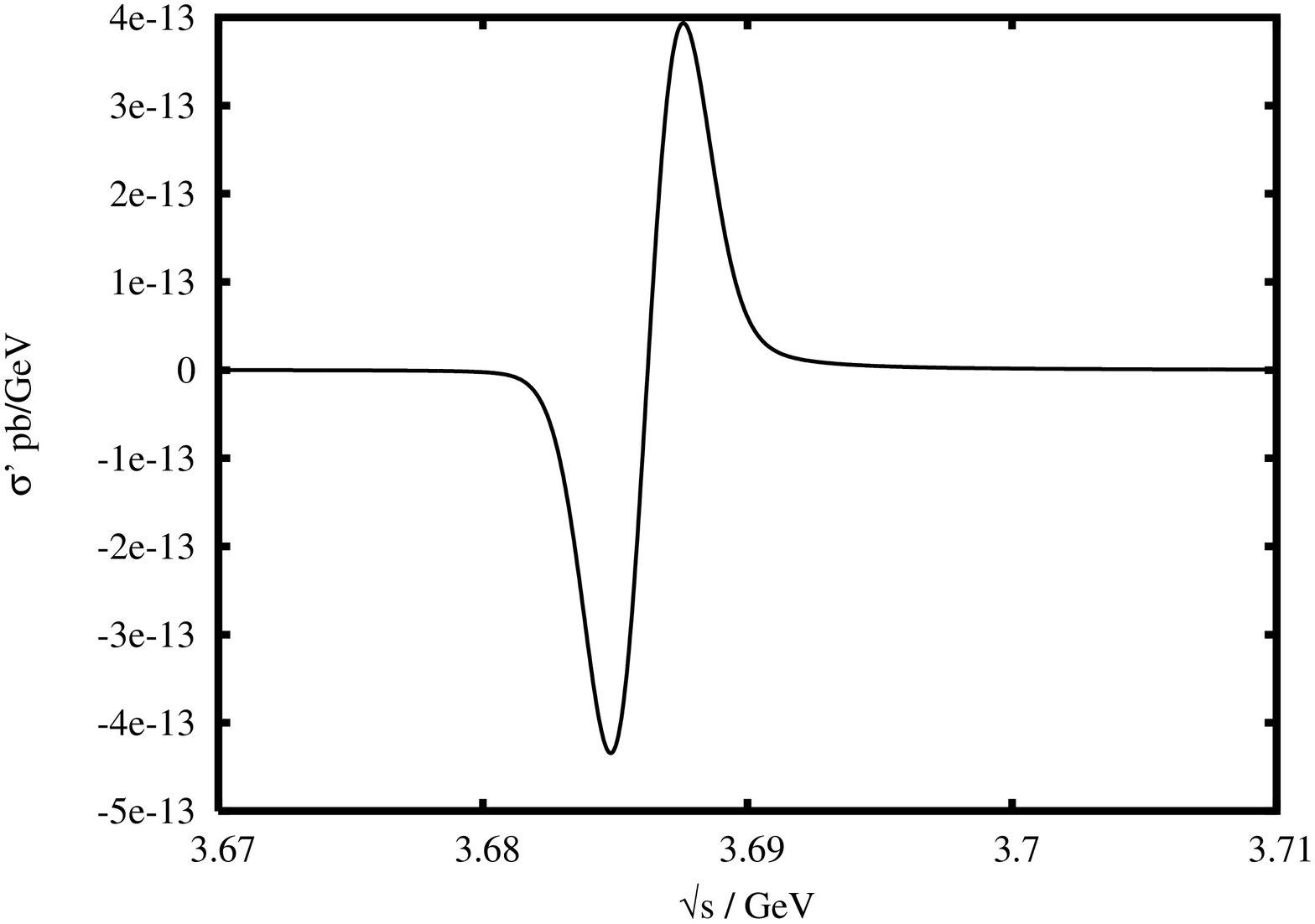}}
\end{minipage}
\caption{The derivative distributions for different parts of the cross
  section. (a) -- (d) is for $\phi = 90^{\circ}$ and (e) -- (h) is for
  $\phi = 180^{\circ}$. (a)(e): $\sigma_{exp}$. (b)(f):
  $\sigma_{exp}^R$. (c)(g): $\sigma_{exp}^{I1}$. (d)(h):
  $\sigma_{exp}^{I2}$. }
  \label{fig:deriv:mass}
\end{figure}

To understand the heterogeneous behavior of curve for $\phi =
180^{\circ}$ from the other ones, we take the curve for $\phi=
90^{\circ}$ as a representative, and show the first order derivatives
of cross sections in \Fig{fig:deriv:mass}, where the left and right
rows correspond to $90^{\circ}$ and $180^{\circ}$, respectively.  The
total cross section $\sigma_{exp}$ is divided into three parts,
i.e. $\sigma_{exp}^R$, $\sigma_{exp}^{I1}$, and $\sigma_{exp}^{I2}$,
as we did in Ref.~\cite{wangbq10}.  The corresponding derivatives of
them are shown sequently in \Fig{fig:deriv:mass}.  Investigation of
those figures indicates that the crucial role for the different
behavior between $90^{\circ}$- and $180^{\circ}$-curves is played by
the $\sigma_{exp}^{I1}$, the derivative variation of which is opposite
to each other.  Furthermore, if we scrutinize the equations in Section
3 of Ref.~\cite{wangbq10}, the sign of the derivative of
$\sigma_{exp}^{I1}$ is determined by the coefficient $A_2 = 6
(\Gamma_{ee}/\alpha)~\cdot~(1~+~{\cal C}~\cos\phi)$, which is positive
when $\phi = 90^{\circ}$ and negative when $\phi = 180^{\circ}$ for
${\cal C}$ = 2.5. The switching point is at $\phi = \cos^{-1}(-1/{\cal
  C}) \approx 113.5^{\circ}$.

\begin{figure}
  \centering
  \subfigure[]{\includegraphics[angle=-90, width=0.4\textwidth]{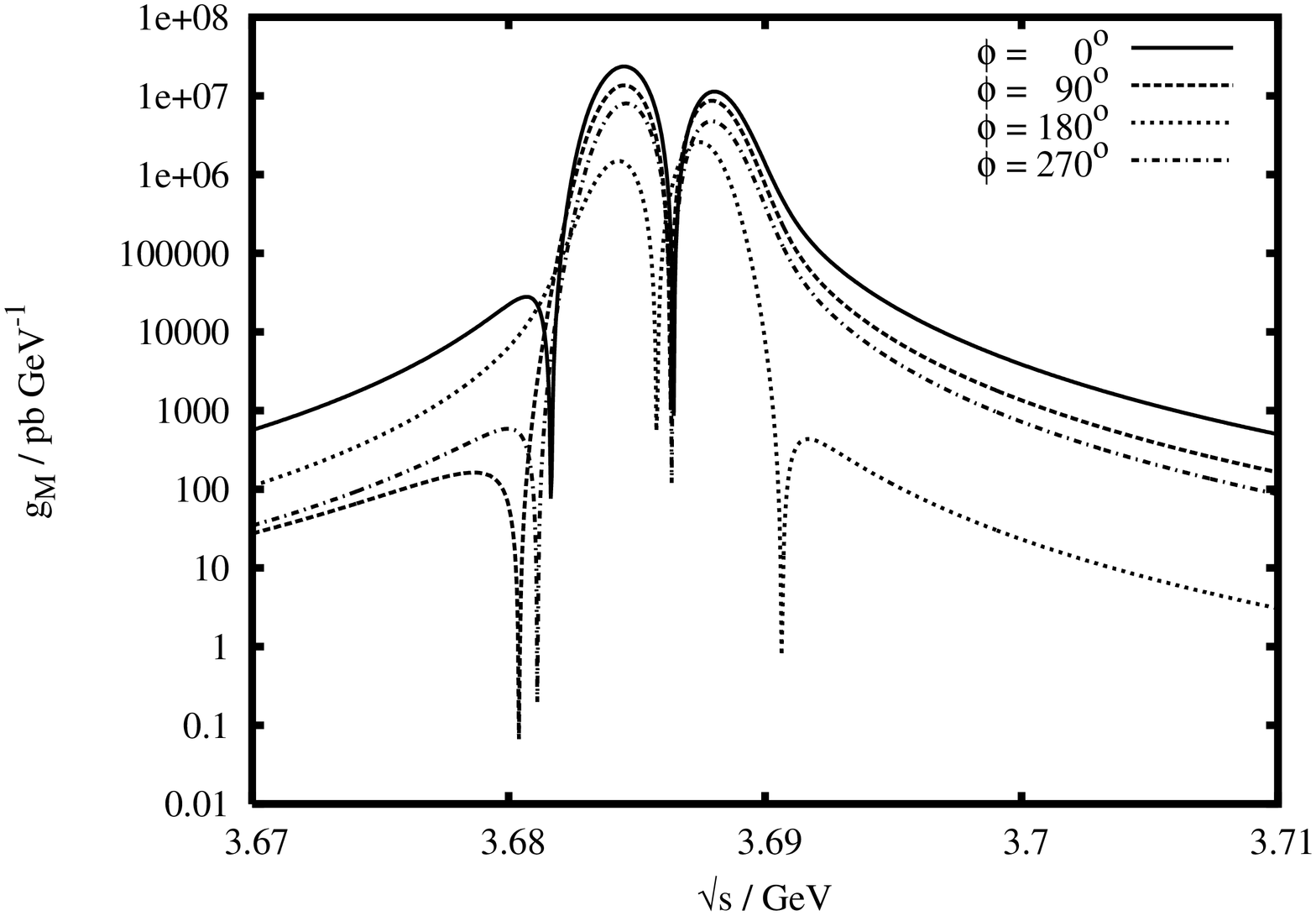}}
  \subfigure[]{\includegraphics[angle=-90, width=0.4\textwidth]{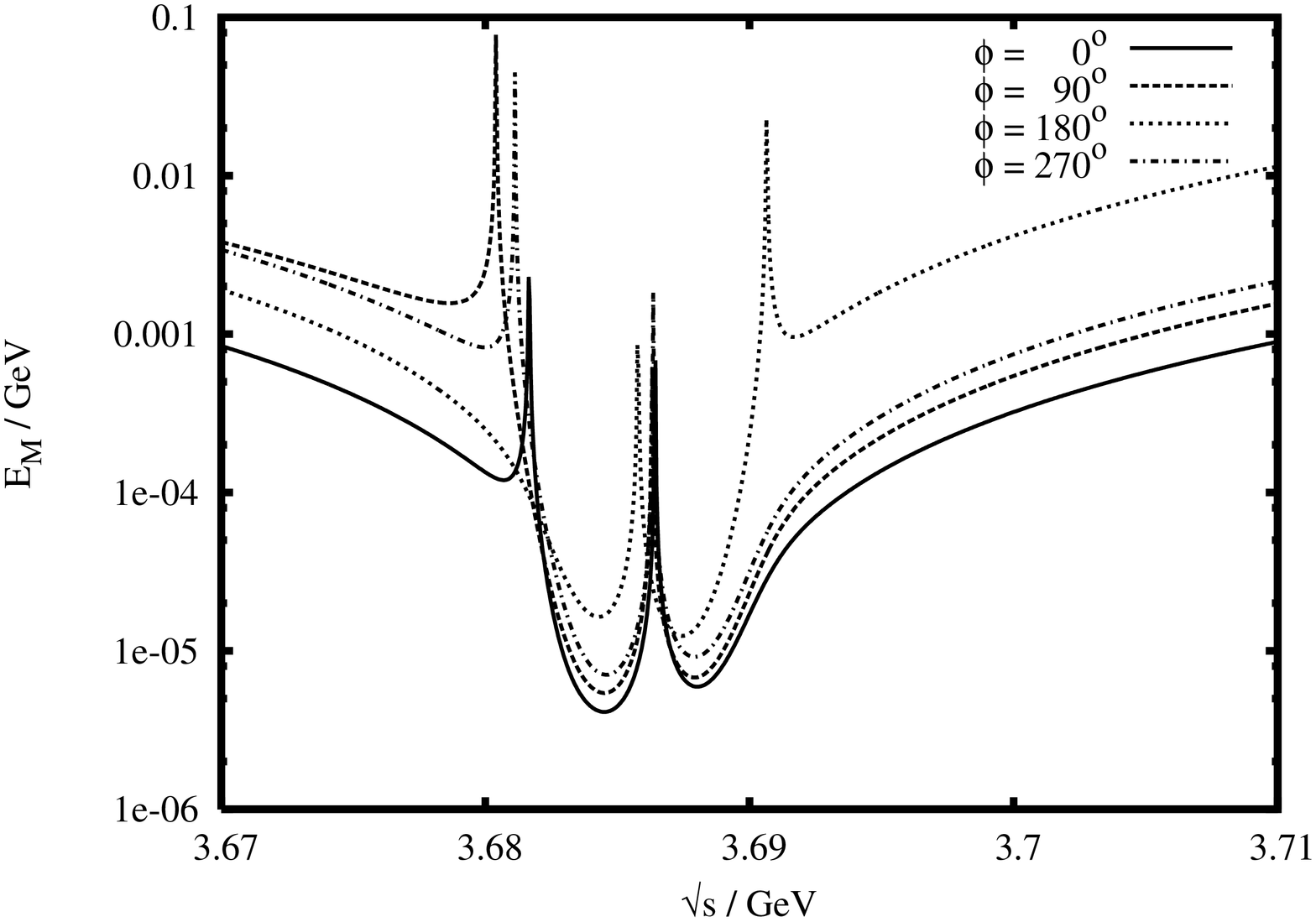}}
  \caption{The results of analyzing $\psip$ mass for relative phase
    $\phi = 0^{\circ}, 90^{\circ}, 180^{\circ}$ and $270^{\circ}$. (a)
    is the $g_M$ value and (b) is the fitting error E$_M$.}
  \label{fig:err:mass}
\end{figure}

\subsubsection{$\psip$ Total Width}

\begin{figure}
  \centering
  \subfigure[]{\includegraphics[angle=-90, width=0.4\textwidth]{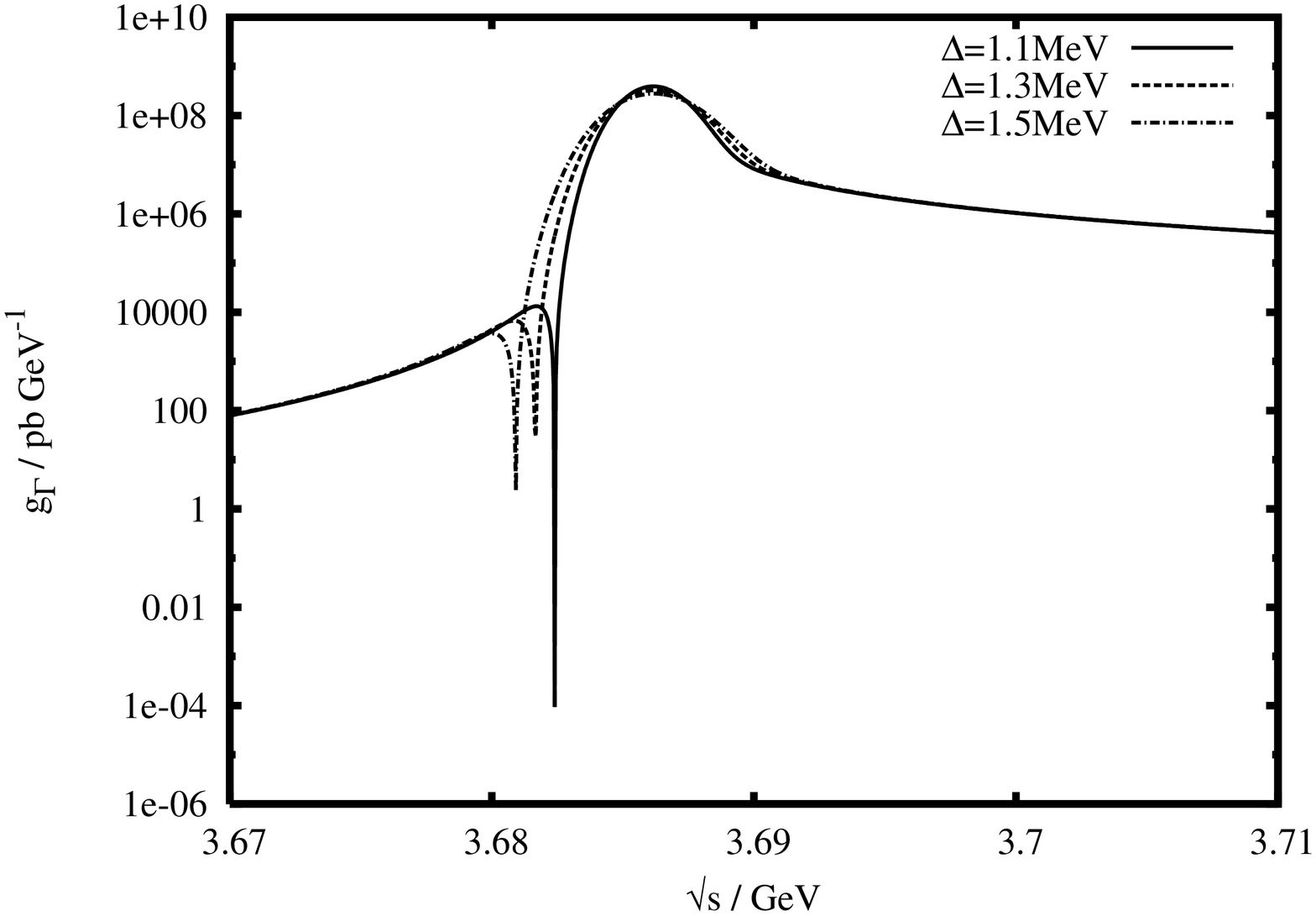}}
  \subfigure[]{\includegraphics[angle=-90, width=0.4\textwidth]{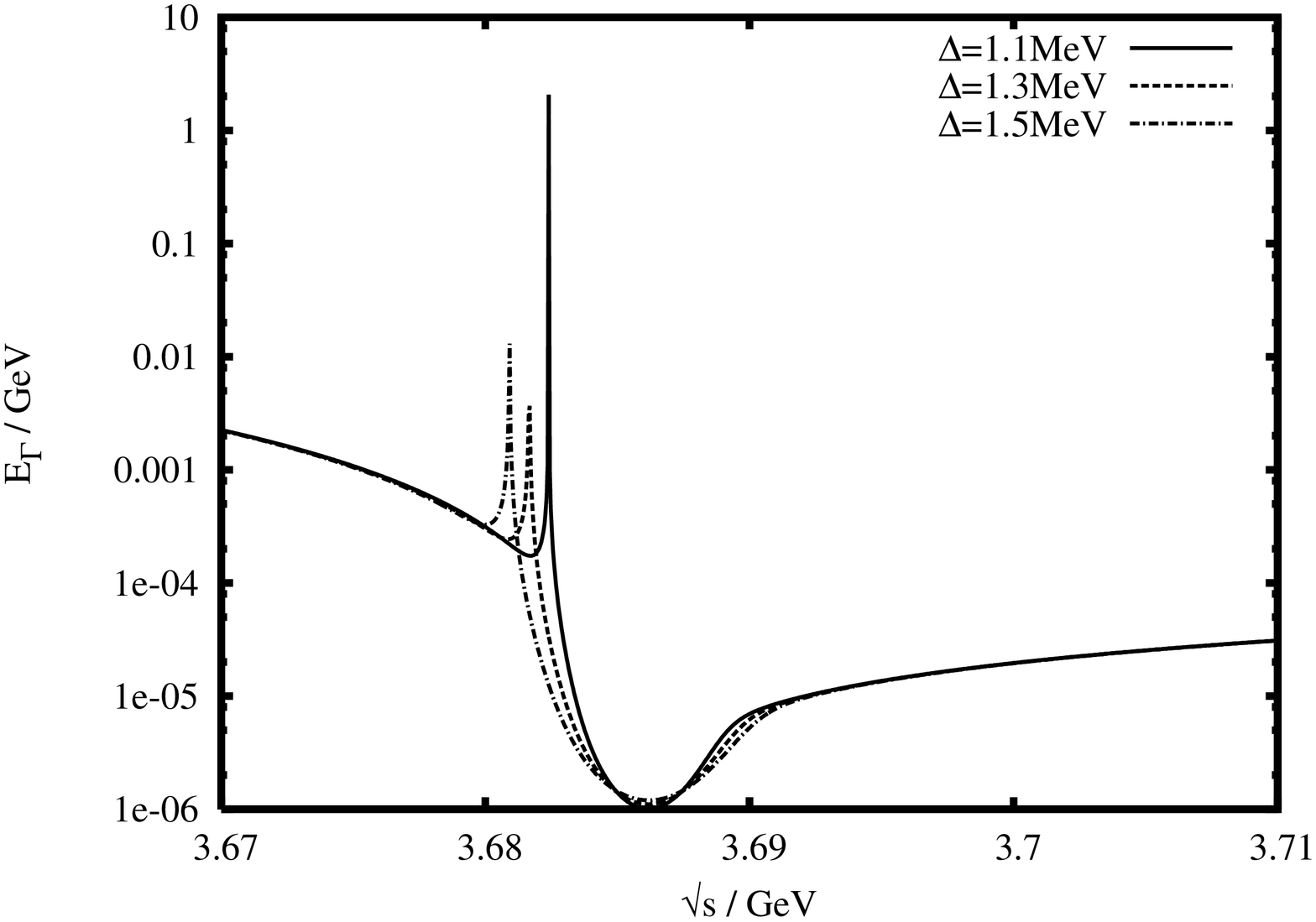}}
  \caption{The results of analyzing $\psip$ total width for $\Delta$ =
    1.1 MeV, 1.3 MeV and 1.5 MeV. (a) is the $g_{\Gamma}$ value and (b) is the
    fitting error E$_{\Gamma}$.}
  \label{fig:err:gamt}
\end{figure}

The optimization results for the total width of $\psip$ resonance
are shown in \Fig{fig:err:gamt}, where three curves corresponding to
distinctive beam energy spreads, $\Delta=1.1, 1.3, 1.5$ MeV, are presented.
With the enhancement of $\Delta$, the position of minimum error, $E_{\Gamma}$,
shifts a little bit rightward along the abscissa. For all circumstances,
the energy position for the maximum $E_{\Gamma}$ is near
3.68 GeV while for the minimum error is around 3.686 GeV.

\section{Conclusion and Discussion}
\label{discussion}

In this paper, one-parameter-and-one-point conclusion is demonstrated
through a theoretical analysis of minimization process instead of the
sampling simulation as we did before. As far as the phase study is
concerned, for the $\psip \to \kk$ process, the optimal data taking
point is determined to at 3.6868 GeV which is near the peak of $\psip$
nominal mass.  The same method is also used to acquire the optimal
point for other resonance parameters, such as the mass and the total
width of $\psip$.

In principle, the idea put forth in Section~\ref{minimize} could be
extended for multi-parameter optimization. Formally speaking, the
vector and matrix quantities, would be adopted for the corresponding
analysis, say, the second order derivative of one parameter are to be
replaced by Hessian matrix, a matrix of second order derivative for
all parameters.

However, there are some problems not easily to be settled. The most
prominent one is how to define ``optimal''. In one parameter scenario,
the optimal data taking point is the one which could make the fitting
error of the parameter reaches its minimum. But for multi-parameters,
there are many options: the sum of relative fitting errors of all
parameters reaches its minimum; the merely fitting error of one major
parameter reaches its minimum while others do not. Different options
lead to distinctive results.  All this makes the situation more
complicated and is left to the study in the further.

\section*{Acknowledgments}

This work is supported by National Natural Science Foundation of China
(11175187, 10825524, 10835001, 10935008), Major State Basic Research
Development Program (2009CB825200, 2009CB825203, 2009CB825206), and
Knowledge Innovation Project of The Chinese Academy of Sciences
(KJCX2-YW-N29).

\appendix

\section{Sampling Technique Methodology}

Suppose there are $N_{pt}$ data taking points in experiment, and the
theoretical number of events in $i$-th energy point could be
calculated as
\begin{equation}
N_i^{the} = L_i\cdot\sigma_i\cdot\varepsilon, i=1,2,\ldots,N_{pt},
\end{equation}
where $L_i$ is the integrated luminosity in $i$-th energy point,
$\varepsilon_i$ is the event selection efficiency.

In sampling technique, the experimentally observed number of events
and its error could be taken as
\begin{equation}
\Delta N_i^{obs} = (N_i^{the})^{1/2},
\end{equation}
\begin{equation}
N_i^{obs} = N_i^{the} + \xi\cdot\Delta N_i^{obs},
\end{equation}
where $\xi$ is a random number which satisfy Gaussian distribution.

Using the observed event number and its error calculated above, the
parameter we interested in (in this paper, the relative phase) could
be fitted by finding the minimum of \Eq{chi2}.

By repeating this process, a large number of observed event number and
error could be generated and so does the fitting parameter and its
error. We can compare these errors with the results obtained from the
method we just developed.

\begin{figure}
\centering
\subfigure[]{\includegraphics[angle=-90, width=0.4\textwidth]{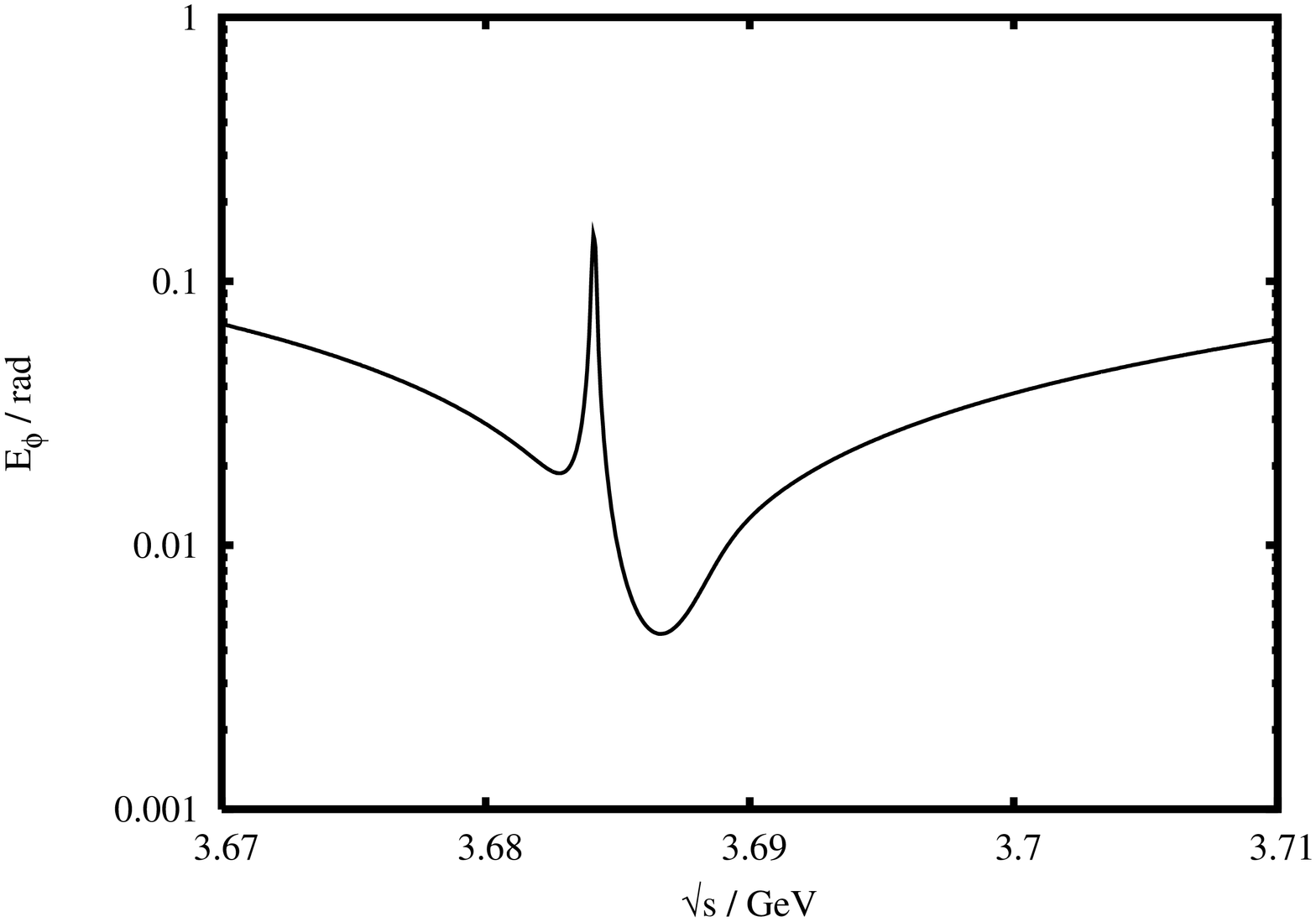}}
\subfigure[]{\includegraphics[angle=-90, width=0.4\textwidth]{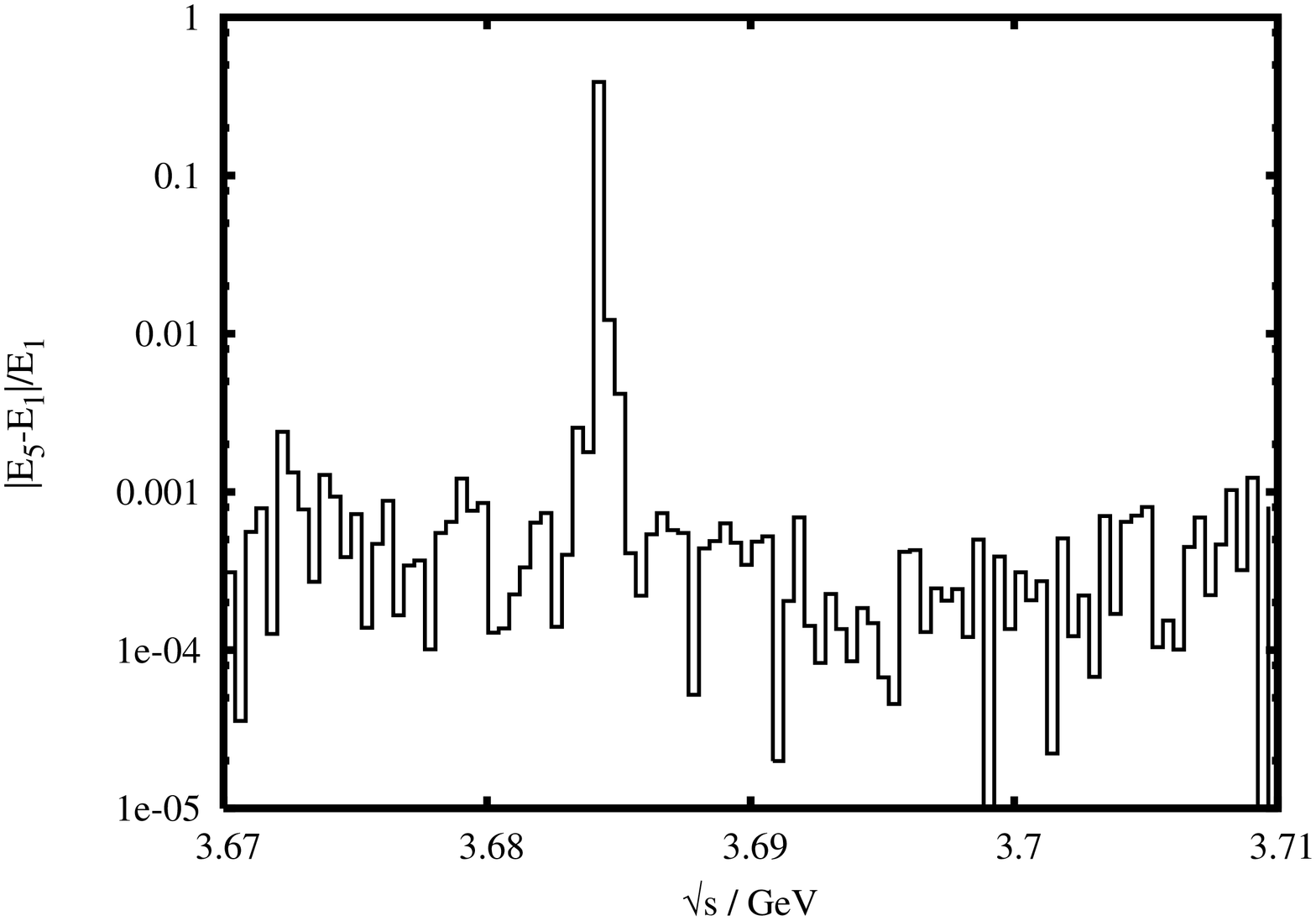}}
\caption{(a) Scan error distribution for five consecutive points; (b)
  The relative difference between five consecutive points and one
  point scan.}
\label{scan-err-diff}
\end{figure}

In section \ref{sec:par:phase}, we scan through the energy region
using one data taking point. To check this result, five consecutive
energy points are used in energy scan. The fitting error of five
consecutive points scan versus the central energy point is shown in
\Fig{scan-err-diff}. This result is similar with one point scan. Their
difference is also shown in \Fig{scan-err-diff}. The difference between
the two scan schemes is generally at the level of one per mille except for
the points around 3.684 GeV, where the variation of error curve is rather rapidly.
As to the five-point scheme, there is at least one point within the region with
comparatively large error.

\end{document}